\documentclass[fleqn,twoside]{article}
\usepackage{gc,amssymb,cite}

\newcommand{\be}{\beq}
\newcommand{\ee}{\eeq}

\newcommand{\R}{{\mathbb R}}

\begin{document}
\twocolumn[

\Title
  {Non-singular solutions in multidimensional cosmology\yy
    with a perfect fluid:  acceleration and variation of $G$}

\Aunames{V.D. Ivashchuk\auth{1,a,b}, S.A. Kononogov\auth{2,a},
         V.N. Melnikov\auth{3,a,b} and M. Novello \auth{4,c}}

\Addresses{
 \addr a {Centre for Gravitation and Fundamental Metrology,
     VNIIMS, 46 Ozyornaya St., Moscow 119361, Russia}
 \addr b {Institute of Gravitation and Cosmology,
     Peoples' Friendship University of Russia,
    6 Miklukho-Maklaya St., Moscow 117198, Russia}
 \addr c { ICRA-Brasil, CBPF, Rua Dr. Xavier Sigaud 150,
     Rio de Janeiro, Brazil}  }


\Abstract {Exact solutions with an exponential behaviour of the scale
 factors are considered in a multidimensional cosmological model describing
 the dynamics of $n+1$ Ricci-flat factor spaces $M_i$ in the presence of a
 one-component perfect fluid. The pressures in all spaces are proportional
 to the density: $p_{i} = w_i \rho$, $i = 0,...,n$. Solutions with
 accelerated expansion of our 3-space $M_0$ and a small enough variation of
 the gravitational constant $G$ are found. These solutions exist for two
 branches of the parameter $w_0$. The first branch describes superstiff
 matter with $w_0 > 1$, the second one may contain phantom matter with
 $w_0 < - 1$. }


]
\email 1 {rusgs@phys.msu.ru}
\email 2 {kononogov@vniims.ru}
\email 3 {melnikov@phys.msu.ru}
\email 4 {novello@cbpf.br}

\section{Introduction}

Fundamental physical constants, relations between them and their possible
variations are a reflection of the situation with unification
 \cite{Solv,KM,AIKM,Paris,1,6}.

Here we are mainly interested in the gravitational constant and its possible
variations. The oldest problem is that of possible temporal variation of
$G$, which arose due to papers by Milne (1935) and Dirac (1937). In Russia,
these ideas were developed in the 60s and 70s by K.P. Staniukovich
 \cite{S,1}, who was the first to consider simultaneous variations of several
fundamental constants.

Our first calculations based on general relativity with a perfect fluid and
a conformal scalar field \cite{ZM} gave $\dot{G}/G$ at the level of
$10^{-11} - 10^{-13}$ per year. Our calculations in string-like \cite{IM1}
and multidimensional models with a perfect fluid \cite{BIM} gave the level
$10^{-12}$, those based on a general class of scalar-tensor theories
 \cite{BMN} and a simple multidimensional model with p-branes \cite{IMW,M-G}
gave for the present values of cosmological parameters $10^{-13}-10^{-14}$
and $10^{-13}$ per year, respectively. Similar estimations were made by
Miyazaki within Machian theories \cite{MZ} giving for $\dot{G}/G$ the
estimate $10^{-13}$ per year and by Fujii --- on the level $10^{-14}-
10^{-15}$ per year \cite{FUJ}. Analysis of one more multidimensional model
with two curvatures in different factor spaces gave an estimate on the level
$10^{-12}$ \cite{DIKM}. Here we continue our studies of variation of $G$ in
one more multidimensional cosmological model with perfect fluid.

\section{The model}

We consider a cosmological model describing the dynamics of $n$ Ricci-flat
spaces in the presence of a one-component ``perfect-fluid'' matter
 \cite{IM5}. The metric of the model
 \beq \label{2.1}
      g= - \exp[2{\gamma}(t)]dt \otimes dt +
      \sum_{i=0}^{n} \exp[2{x^{i}}(t)] g^{i}
 \eeq
is defined on the manifold
 \beq \label{2.2}
      M = \R \times M_{0} \times \ldots \times M_{n},
 \eeq
where $M_{i}$ with the metric $g^{i}$ is a Ricci-flat space
of dimension $d_i$, $i = 0, \ldots ,n $; $n \geq 2$.
The multidimensional Hilbert-Einstein equations have the form
 \be \label{2.3}
       R^{M}_{N}-\frac{1}{2}\delta^{M}_{N}R = \kappa^{2}T^{M}_{N},
 \ee
where $\kappa^{2}$ is the gravitational constant, and the
energy-momentum tensor is adopted as
 \beq                                           \label{2.6}
    (T^{M}_{N}) = \diag (- \rho,\  p_{0} \delta^{m_{0}}_{k_{0}},
    \ldots ,\ p_{n} \delta^{m_{n}}_{k_{n}}),
 \eeq
describing, in general, an anisotropic fluid.

We assume the pressures of this ``perfect'' fluid in all spaces to be
proportional to the density,
 \beq \label{2.8}
      p_i(t) = (1- u_i/d_i) \rho(t),
 \eeq
where $u_i = \const$, $i = 0, \ldots ,n$. We also put $\rho \neq 0$.

We also impose the following restriction on the vector $u = (u_i) \in
 \R^{n+1}$:
 \be \label{2.9}
        \aver{u,u}_{*} \neq 0.
 \ee
Here, the bilinear form $\aver {.,.}_{*}: \R^{n+1} \times \R^{n+1}
 \to \R$ is defined by the relation \be             \label{2.10}
        \aver {u,v}_{*} = G^{ij} u_i v_j,
 \ee
 $u,v \in \R^{n+1}$, where
\beq                \label{2.11}
     G^{ij} = \frac{\delta^{ij}}{d_i} + \frac{1}{2-D}
\eeq
 are components of the matrix inverse to the matrix of
 the minisuperspace metric \cite{IM2,IMZ}
\beq \label{2.12}
      G_{ij} = d_{i} \delta_{ij} - d_{i} d_{j}.
\eeq
 In (\ref{2.11}), $D = 1 + \sum_{i=0}^{n} d_i$ is the total
 dimension of the manifold $M$ (\ref{2.2}).

  The restriction (\ref{2.9}) reads
\beq  \label{2.13}
   \aver{u,u}_{*}  =   \sum_{i= 0}^{n} \frac{(u_{i})^{2}}{d_{i}}
   + \frac{1}{2-D}\biggl(\sum_{i= 0}^{n} u_{i}\biggr)^{2} \neq 0.
\eeq

 \section{Solutions with exponential scale factors}

Here, we consider a special family of solutions with an exponential
behaviour of the scale factors from \cite{IM5,IM10} with the metric written
in the synchronous time parametrization
\be \label{3.1}
     g=- dt_s \otimes dt_s + \sum_{i= 0}^{n} a_i^2(t_s) g^{i}.
\ee
 Solutions with a an exponential behaviour  of the scale factors
 take place for
\be   \label{3.2}
     \aver {u^{(\Lambda)} - u, u}_{*} = 0.
\ee
 Here and below the vector
\be  \label{3.3}
         u^{(\Lambda)}_i = 2d_i
\ee
 corresponds to the $\Lambda$-term ``fluid'' with $p_i = - \rho$ (vacuum-like
 matter).

 In this case, the solutions are determined by the metric (\ref{3.1})
 with the scale factors
\be   \label{3.4}
       a_i = {a_i}(t_s)  =   A_i \exp(\nu^i t_s),
\ee
and the density
\be   \label{3.4a}
  \rho = \const.
\ee
  Here
\be \label{3.5}
     \nu^i  = \eps u^i
     \sqrt{- \frac{2 \kappa^2 \rho}{\aver{u,u}_{*}}},
\ee
 where $\eps = \pm 1$, $u^i =   G^{ij} u_i$ and $A_i$ are
 positive constants, $i = 0, \dots, n$. Here
 $\rho > 0$  for $\aver{u,u}_{*} < 0 $ and
 $\rho < 0$  for $\aver{u,u}_{*} > 0 $.

 The model under consideration was integrated in \cite{IM5} for
 $\aver{u,u}_{*} < 0$. The solutions from \cite{IM5} were generalized in
 \cite{IM10} to the case when a massless minimally coupled scalar field was
 added. Families of exceptional solutions with power-law and exponential
 behaviours of the scale factors in terms of synchronous time were singled
 out in \cite{IM10} and correspond to a constant value of the scalar field:
 $\varphi = \const$. When the scalar field is omitted, we are led to
 solutions presented in \cite{AIKM} and above for power-law and exponential
 cases, respectively (in \cite{IM5} these solutions were originally written
 in the harmonic time parametrization). It may be verified that the
 exceptional solutions with an exponential dependence of the scale factors
 are also valid when the restriction $\aver{u,u}_{*} < 0$ is replaced by
 (\ref{2.9}).

 \section{Acceleration and variation of $G$}

 In this section, the metric $g^0$ is assumed to be flat, and $d_0 =3$. The
 subspace $(M_0,g^0)$ describes ``our'' 3-dimensional space and
 $(M_i,g^i)$ the internal factor spaces.

 We are interested in solutions with accelerated expansion of our space and
 small enough variations of the gravitational constant obeying the present
 experimental constraints, see \cite{IMW}:
\be \label{4.1v}
        |\dot{G}/(GH)|(t_{s0}) < 0.1,
\ee
 where
\be \label{4.1h}
                H = \frac{\dot{a_0}}{a_0}
\ee
 is the Hubble parameter. We suppose that the internal spaces are
 compact. Hence our 4-dimensional constant is (see \cite{BIM})
\be \label{4.1g}
         G  = \const \cdot \prod\nolimits_{i=1}^{n}( a_{i}^{-d_i}).
\ee

  We will use the following explicit formulae for the contravariant
  components:
\be \label{4.2}
   u^i = G^{ij} u_j = \frac{u_i}{d_i} + \frac{1}{2-D} \sum_{j =0}^{n} u_j,
\ee
  and the scalar product reads
\bearr \label{4.3}
   \aver{ u^{(\Lambda)} - u, u }_{*}
\nnn \nq
   =  - \sum_{i= 0}^{n} \frac{(u_{i})^{2}}{d_i}
      - \frac{2}{D-2}\sum_{i = 0}^n u_i
      + \frac{1}{D-2}\Biggl(\sum_{i= 0}^{n} u_{i}\Biggr)^2.
\ear

\subsection{Exponential expansion with acceleration}

 For solutions under consideration, an accelerated
 expansion of our space takes place for
\be \label{4.4}
     \nu^0 > 0.
\ee
  {\bf Remark:  D= 4 case.}
 {\em For  $D=4$, when the internal spaces are absent, we get
  $u^0 = - u_0/6$ and
\bearr                                   \label{4.2pa}
    \aver { u, u }_{*} = - \frac {1}{6} u_0^2,
         \\ \lal    \label{4.2p}
    \aver {u^{(\Lambda)} -u, u }_{*} = \frac {1}{6}(u_0 -6) u_0 = 0,
\ear
  which implies  $u_0 =  6$,   or, equivalently,
\be     \label{4.3a}
        p =  - \rho.
\ee
 We get
\be \label{4.4a}
      \nu^0 = - \eps   \sqrt{ \frac{\kappa^2 \rho}{3}},
 \ee
  which agrees with the well-known result for $D =4$: the de-Sitter solution
  with the cosmological constant $\Lambda = \kappa^2 \rho > 0$.
  The condition  $\nu^0 >0$ is equivalent to $\eps = -1$.}

  For our exponential solutions we get
\be \label{4.5a}
     \frac{\dot{G}}{G} =  - \sum_{j =1}^{n} \nu^i d_i, \cm
     H = \frac{\dot{a_0}}{a_0} = \nu^0,
\ee
  and hence
\be \label{4.5}
   \dot{G}/(GH) =- \frac{1}{\nu^0} \sum_{j =1}^{n} \nu^i d_i \equiv \delta.
\ee
  The constant parameter $\delta$ describes variation of the gravitational
  constant and, according to  (\ref{4.1v}),
\be \label{4.6}
            |\delta| < 0.1.
\ee

  It follows from the definition of $\nu^i$ in (\ref{3.5}) that
\be \label{4.7}
        \delta = - \frac{1}{u^0} \sum_{i =1}^{n} u^i d_i,
\ee
   or, in terms of covariant components (see (\ref{4.2})),
\be \label{4.8}
    \delta = - \frac{(D-4) u_0 -2 \sum_{i =1}^{n} u_i}{\frac{1}{3} (5 - D)
                u_0 + \sum_{i =1}^{n} u_i}.
\ee

   Thus the relations  (\ref{4.3}), (\ref{4.4}), (\ref{4.6}), (\ref{4.8})
   and the constraint (\ref{3.2}) determine a set of parameters $u_i$
   compatible with the acceleration and tests on $G$-dot.

   In what follows we will show that these relations do really determine a
   non-empty set of parameters $u_i$ describing the equations of state.

\subsubsection{The case of constant $G$}

   Consider the most important case $\delta =0$, i.e., when the variation
   of $G$ is absent: $\dot{G} = 0$.

   Indeed, there is a tendency of lowering the upper bound on $\dot G$.
   Moreover, according to arguments of \cite{BZhuk}, $\delta < 10^{-4}$.
   This severe constraint just follows from the identity
\be \label{4.Al}
       \dot{G}/G = \dot{\alpha}/\alpha
\ee
   that takes place in some multidimensional models. Here $\alpha$ is the fine
   structure constant.

\medskip\noi
 {\bf Isotropic case.} We first consider the isotropic case when the
 pressures coincide in all internal spaces.  This takes place when
\be  \label{4.9}
        u_i = v d_i, \cm  i = 1, \ldots, n.
\ee
 For pressures in internal spaces we get from  (\ref{2.8})
\be \label{4.10}
      p_i = (1- v) \rho,  \cm i = 1, \ldots, n .
\ee
   Then we get from (\ref{2.13}) and (\ref{4.3})
\bearr   \label{4.11a} \nhq                                  
    \aver{u, u}_{*} = \frac{1}{2-D}
        [- \fract{1}{3} (d-1) u_0 + 2d u_0 v- 2d v^2],
\\ \lal                         \label{4.11b}
    \aver{ u^{(\Lambda)} - u, u }_{*} =
                    \frac{1}{2-D}\bigl[ 2 u_0 + 2dv
\nnn  \cm  \cm
    + \fract{1}{3}(d-1) u_0^2- 2d u_0 v+ 2d v^2 \bigr].
\ear
  Here and in what follows we denote $d = D-4$.

  For $\delta = 0$, we get in the isotropic case
\be  \label{4.12}
        v = u_0/2,
\ee
   or, in terms of pressures,
\be  \label{4.12p}
          p_i= ( 3 p_0 - \rho)/2, \cm  i = 1, \ldots, n.
\ee

   Substituting  (\ref{4.12}) into (\ref{4.11a}) and  (\ref{4.11b}), we get
\bearr   \label{4.13a}
     \aver{u, u}_{*} = - u_0^2/6,
\\ \lal                     \label{4.13b}
    \aver{u^{(\Lambda)} - u, u }_{*} = u_0 (u_0 -6)/6.
\ear

   Here, we obtain the same relations as for $D = 4$. For our solution, we
   should put $u_0 \neq 0$ and hence, due to (\ref{3.2}),
\be  \label{4.12pp}
           u_0 = 6,
\ee
   i.e.,
\be  \label{4.12o}
          p_0= - \rho, \cm  p_i = - 2 \rho,  \cm   i > 0.
\ee

   Using (\ref{4.9}) and (\ref{4.12}), we get  $u^0 = - u_0/6 = -1$ and $u^i
   = 0$  for $i > 0$,  hence  $ \nu_i =0$ for $i = 1, \ldots, n$, i.e., all
   internal spaces are static.

   The metric  (\ref{3.1}) reads in our case
\be \label{4.14}
        g=- dt_s \otimes dt_s + A_0^2 \exp (2 \nu^0 t_s) g^{0} +
                \sum_{i= 1}^{n} A_i^2 g^{i},
\ee
   where $A_i$ are positive constants, and
\be   \label{4.15}
   \nu^0 = - \eps \sqrt{\frac{\kappa^2 \rho}{3}}.             
\ee
   For accelerated expansion we get $\eps = -1$.
   We see that the power $\nu^0$ is the same as in case $D =4$.

\medskip\noi
 {\bf Anisotropic case.} Consider the anisotropic (w.r.t. internal spaces)
 case with $\delta = 0$, or, equivalently (see (\ref{4.8})),
\be  \label{4.18a}
        (D-4) u_0 = 2 \sum_{i =1}^{n} u_i.
\ee
   This implies
 \bearr  \label{4.18}
     \aver{ u^{(\Lambda)} - u, u }_{*} = \fract{1}{6}u_0 (u_0 -6) - \Delta,
 \\ \lal   \label{4.18b}
     \aver{ u, u }_{*} = - \fract{1}{6}  u_0^2 + \Delta,
\ear
    where
\be \label{4.19}
  \Delta= \sum_{i=1}^n \frac{u_i^2}{d_i}
   - \frac{1}{d} \biggl(\sum_{i=1}^n  u_i\biggr)^2 \geq 0,  \quad\
        d = D - 4.
\ee

 The inequality in (\ref{4.19}) can be readily proved
 using  the well-known  Cauchy-Schwarz inequality:
 \be \label{4.19CS}
   \biggl(\sum_{i=1}^n b_i^2\biggr) \biggl(\sum_{i=1}^n c_i^2\biggr)
        \geq \biggl(\sum_{i=1}^n  b_i c_i \biggr)^2.
\ee
  Indeed, substituting $b_i = \sqrt{d_i}$ and  $c_i = u_i/ \sqrt{d_i}$
  into  (\ref{4.19CS}), we get  (\ref{4.19}). The equality in (\ref{4.19CS})
  takes place only when the vectors $(b_i)$ and $(c_i)$ are linearly
  dependent, which for our choice reads:  $u_i/ \sqrt{d_i} = v \sqrt{d_i}$
  where $v$ is constant. Thus $\Delta = 0$ only in the isotropic case
  (\ref{4.9}). In the anisotropic case we get $\Delta > 0$.

  In what follows we will use the relation
\be \label{4.19a}
      \aver { u^{(\Lambda)} - u, u }_{*} =
         \fract{1}{6} (u_0 - u_0^{+})(u_0 - u_0^{-}),
\ee
   where
\be \label{4.19b}
       u_0^{\pm} = 3 \pm \sqrt{9 + 6 \Delta}
\ee
  are roots of the quadratic trinomial (\ref{4.18}) obeying
\be \label{4.19c}
        u_0^{-} < 0, \qquad   u_0^{+} > 6 \qquad
            {\rm for}\quad \Delta > 0.
\ee
    It follows from  (\ref{4.18a}) that $u^0 = - u_0/6$ and hence
\be  \label{4.20}
   \nu^0 = - \eps \frac{u_0}{6} \sqrt{\frac{12 \kappa^2 \rho}{
    u_0^2 - 6\Delta}}.
\ee
  Here  (see (\ref{3.2}))
 \be  \label{4.20a}
   u_0 = u_0^{\pm}.
 \ee

  Accelerated expansion of our space takes place
  when $\nu^0 > 0$, or, equivalently, when either
\bearr  \label{4.22a}
    {\bf (A)} \quad u_0 =  u_0^{-}, \quad \eps = 1  \quad {\rm or}
\\ \lal       \label{4.22b}
{\bf (B)} \quad u_0 =  u_0^{+}, \quad \eps = -1.
\ear

  In terms of the parameter $w_0$,
\be  \label{4.24}
      p_0 = w_0 \rho, \qquad   w_0 = 1 - u_0/3,
\ee
   these two branches read:
\bearr \nq
   {\bf (A)}  \label{4.25a} \cm
   w_0 = w_0^{-} = \sqrt{1 + \fract{2}{3} \Delta},
\yyy \nq
   {\bf (B)}   \label{4.25b} \cm
   w_0 = w_0^{+} = - \sqrt{1 + \fract{2}{3} \Delta}.
\ear
     The first branch (A) describes super-stiff matter ($w_0 > 1$) with
     negative density $\rho < 0$.

     The second  branch (B) corresponds to matter with positive density
     (since $\aver{u, u }_{*} < 0$). This matter is phantom (i.e., $w_0
     < - 1$) when $\Delta > 0$.

\subsubsection{The case of varying $G$}

   Now we consider the case $\delta \neq 0$, i.e., when $\dot{G} \neq 0$. In
   what follows, we use the observational bound (\ref{4.6}):  $|\delta| <
   0.1$,  stating the smallness of $\delta$.

   Using (\ref{4.8}), we get
\be  \label{5.18a}
            \sum_{i =1}^{n} u_i = \frac{1}{2} d b u_0 ,
\ee
   where $d = D-4$ and
\be  \label{5.18bd}
         b  = b(\delta) = \frac{1+ \delta (1-d)/(3d)}{1 -  \delta/2}.
\ee
   For the scalar product we get from (\ref{5.18a})
\bearr  \label{5.18}
     \aver{ u^{(\Lambda)} - u, u }_{*} =
     \fract{1}{6} A u_0^2  - B u_0  - \Delta,
 \\ \lal   \label{5.18b}
     \aver { u, u }_{*} = - \fract{1}{6} A u_0^2 + \Delta,
\ear
  where $\Delta$ was defined in (\ref{4.19}) (see (\ref{2.13}) and
  (\ref{4.3})),
\bear  \label{5.18c}
    \frac{A}{6} \eql \frac{1}{d+2}\biggl(1 + \frac{d}{2}b\biggr)^2 -
               \frac{d}{4}b^2 - \frac{1}{3},
\\  \label{5.18d}                                            
    B \eql \frac{1}{d+2}  (2 + db).
\ear
  Using (\ref{5.18bd}), we obtain the explicit formulae
\bear  \label{5.18cc}
    A \eql A(\delta) = 1 - \frac{(d+2) \delta^2}{12d (1 - \delta/2)^2},
\\  \label{5.18dd}
    B \eql B(\delta) = \frac{1 - \delta/3}{1 - \delta/2}.
\ear

   Due to $|\delta| < 0.1$,  $A$ is positive,  $A > 0$, and close to unity:
   $|A -1| < \frac{1}{3} 10^{-2}$.

  For the contravariant component $u^0$ we get from (\ref{4.2}) and
  (\ref{5.18a}):
\be  \label{5.18f}
    u^0 = - C u_0/6,
\ee
  where
\be  \label{5.18ff}
    C = C(\delta) = 3B  - 2  = 1/(1 - \delta/2).
\ee
  It follows from (63) and (\ref{5.18f}) that (see (\ref{3.5}))
\be  \label{5.20}
  \nu^0 = - \eps \frac{C u_0}{6} \sqrt{\frac{12 \kappa^2 \rho}{
   A u_0^2 - 6\Delta}}.
\ee
  Here
  (due to  (\ref{3.2}))
\be \label{5.19b}
    u_0 =  u_0^{\pm}(\delta) = \frac{1}{A}
                  (3 B \pm \sqrt{9 B^2  + 6 A \Delta} )
\ee
  are roots of the quadratic trinomial (\ref{5.18}).

\medskip\noi
  {\bf Isotropic case.} Let us consider the isotropic case (\ref{4.9}).
  Then we obtain from  (\ref{5.18a})
\be  \label{5.12}
        v  = d b u_0 /2,
\ee
  or, in terms of pressures
\be  \label{5.12p}
    p_i= \frac{1}{2} [ 3b p_0 + (2 -3b) \rho], \qquad  i = 1, \ldots, n.
\ee
    For scalar products we get
\bearr   \label{5.13a}
    \aver{ u, u}_{*} = - A u_0^2/6,
\\ \lal              \label{5.13b}
   \aver { u^{(\Lambda)} - u, u }_{*} = u_0 (A u_0 - 6 B)/6.
\ear

    For our solution, we should put $u_0 \neq 0$ and hence $u_0 = 6B/A > 0$.
    The metric  (\ref{3.1}) reads in our case
\be \label{5.14}
        g= - dt_s \otimes dt_s +  A_0^2 \e^{2 \nu^0 t_s} g^{0} +
          \e^{2 \nu t_s} \sum_{i= 1}^{n} A_i^2 g^{i},
\ee
   where $A_i$ are positive constants,
\bear                                                   \label{5.15}
     \nu^0 \eql  - \eps
    \frac{C u_0}{6} \sqrt{\frac{12 \kappa^2 \rho}{A u_0^2}},
    \qquad {\rm and}
\\            \label{5.15a}
   \nu \eql   \nu^i =   \eps \frac{\delta u_0}{6 d(1 - \delta/2)}
     \sqrt{\frac{12 \kappa^2 \rho}{A u_0^2}},
\ear
   $i = 1, \ldots, n$.
   The last formula follows from (\ref{3.5}) and
\be                               \label{5.15b}
   u^i =  \frac{u_0 \delta}{6d(1 - \delta/2)}.
\ee
  We see that the power $\nu^0$ does not coincide, for $\delta \neq 0$, with
  that in case $D =4$.

  The accelerated expansion condition for our 3D space, $\nu^0 > 0$, reads
  in this case
\be \label{5.17}
     u_0 = \frac{6B(\delta)}{A(\delta)}, \qquad  \eps  = -1
\ee
  or, equivalently, in terms of $w_0$  (\ref{4.24}) ($p_0 = w_0 \rho$)
\be \label{5.17a}
   w_0 =  w_{0}^{+}(\delta) =  1 - \frac{2B(\delta)}{A(\delta)}.
\ee

  For $\delta > 0$, we get isotropic contraction of the whole internal
  space $M_{1} \times \ldots \times M_{n}$. In this case,
  $w_{0}^{+}(\delta)  < - 1$, and hence phantom matter occurs with the
  equation of state close to the vacuum one since
\be \label{5.17b}
     w_{0}^{+}(\delta) +  1 =  - \frac{\delta (1 + \delta/d)}
     {3[1 - \delta + (d -1) \delta^2/(6d)]}.
\ee
  For small $\delta$ we have $w_{0}^{+}(\delta) =-1 - \delta/3 + O(\delta^2)$.

  For $\delta < 0$ we obtain isotropic expansion of the whole internal
  space. Then $w_{0}^{+}(\delta) > - 1$, and phantom matter does not occur.

\medskip\noi
  {\bf Anisotropic case.} Now we consider the anisotropic case
  $\Delta > 0$ when $\delta \neq 0$.
  Here $u_0 = u_0^{\pm}(\delta)$, see (\ref{5.19b}).

 Accelerated expansion of our 3-dimensional space takes place
 when $\nu^0 > 0$, or, equivalently, when either
\bearr  \label{5.22a}
    {\bf (A)} \quad u_0 = u_0^{-}(\delta), \qquad  \eps = 1
            \qquad  {\rm or}
\\ \lal    \label{5.22b}
    {\bf (B)} \quad u_0 =  u_0^{+}(\delta), \qquad \eps = - 1 .
\ear

In terms of the parameter $w_0$ ($p_0 = w_0 \rho$,
 $w_0 = 1 - u_0/3$) these two branches read
\bearr  \label{5.25a}
    {\bf (A)} \quad w_0 = w_0^{-}(\delta),
\\ \lal         \label{5.25b}
    {\bf (B)}  \quad w_0 = w_0^{+}(\delta),
\ear
where
\be  \label{5.26a}
    w_0^{\pm}(\delta) = 1 - {u_0^{\pm}(\delta)}/{3}.
\ee
     For small $\delta$ we have
\be  \label{5.27a}
  w_0^{\pm}(\delta) = w_0^{\pm}(0) -   \frac{\delta}{6}
    \biggl(1 \pm \frac{3}{\sqrt{9 + 6 \Delta}}\biggr) + O(\delta^2).
 \ee

  Thus for small $\delta$ the parameter $w_0$ has a small deviation from
  that obtained for $\delta = 0$. For small $\delta$, $w_0$ shifts by
  $O(\delta)$ term.

  The first branch (A) describes superstiff matter, $w_0 > 1$, since
  $w_0^{-}(\delta) > 1$ due to  $u_0^{-}(\delta) < 0$. It may be shown that
  the density is negative in this case since $\aver{u, u }_{*} > 0$.

  For branch (B) we find that
\be  \label{5.28b}
      w_{0}^{+}(\delta) < - 1
\ee
   only if
\be  \label{5.28c}
      \Delta > 6 [A(\delta) -B(\delta)] = - \delta/(1 - \delta/2)^2.
\ee
   This is a condition on the appearance of ``phantom'' matter. For $\delta
   > 0$ this inequality is valid, but for $\delta < 0$ it is satisfied only
   for a big enough value of the anisotropy parameter $\Delta$, see
   (\ref{5.28c}).

 \section{Conclusions}

 We have considered multidimensional cosmological models describing the
 dynamics of $n+1$  Ricci-flat factor spaces $M_i$ in the presence of a
 one-component anisotropic (perfect) fluid with pressures in all spaces
 proportional to the density: $p_{i} = w_i \rho$, $i = 0,...,n$. Solutions
 with accelerated expansion of our 3-dimensional space $M_0$ and small
 enough variation of the gravitational constant $G$ were found.  In the
 general non-isotropic case, these solutions exist for two branches of the
 parameter $w_0$: (A) from (\ref{5.25a}) and (B) from (\ref{5.25b}).
 Branch (A) describes super-stiff matter with $w_0 > 1$ while branch (B)
 may corresponds to phantom matter with $w_0 < - 1$.

\Acknow{The  work of V.D.I. and  V.N.M. was supported in part
   by the  DFG grant  Nr. 436 RUS 113/807/0-1 and also
   by the Russian Foundation for   Basic Research, grant Nr.
   05-02-17478.}

 \end{document}